\documentclass[10pt,aps,prl,twocolumn,superscriptaddress,a4paper]{revtex4}

\usepackage[utf8]{inputenc}
\usepackage{hyperref}
\usepackage{epsfig}
\usepackage{graphicx}
\usepackage{amsmath}
\usepackage{amssymb}
\usepackage{color}

\graphicspath{{Images/}}

\begin{document}

\title{
Non-destructive testing of composite plates by holographic vibrometry
}

\author{Francois Bruno}

\affiliation{
Institut Langevin. Fondation Pierre-Gilles de Gennes. Centre National de la Recherche Scientifique (CNRS) UMR 7587, Institut National de la Sant\'e et de la Recherche M\'edicale (INSERM) U 979, Universit\'e Pierre et Marie Curie (UPMC), Universit\'e Paris 7. Ecole Sup\'erieure de Physique et de Chimie Industrielles (ESPCI) - 1 rue Jussieu. 75005 Paris. France
}

\author{J\'er\^ome Laurent}

\affiliation{
Institut Langevin. Fondation Pierre-Gilles de Gennes. Centre National de la Recherche Scientifique (CNRS) UMR 7587, Institut National de la Sant\'e et de la Recherche M\'edicale (INSERM) U 979, Universit\'e Pierre et Marie Curie (UPMC), Universit\'e Paris 7. Ecole Sup\'erieure de Physique et de Chimie Industrielles (ESPCI) - 1 rue Jussieu. 75005 Paris. France
}

\author{Claire Prada}

\affiliation{
Institut Langevin. Fondation Pierre-Gilles de Gennes. Centre National de la Recherche Scientifique (CNRS) UMR 7587, Institut National de la Sant\'e et de la Recherche M\'edicale (INSERM) U 979, Universit\'e Pierre et Marie Curie (UPMC), Universit\'e Paris 7. Ecole Sup\'erieure de Physique et de Chimie Industrielles (ESPCI) - 1 rue Jussieu. 75005 Paris. France
}

\author{Benjamin Lamboul}

\affiliation{
Composite Structures and Materials Department, Office National d'\'Etudes et de Recherches A\'eronautiques (ONERA), BP72 - 29 avenue de la Division Leclerc, 92322 Chatillon Cedex. France
}

\author{Bruno Passilly}

\affiliation{
Composite Structures and Materials Department, Office National d'\'Etudes et de Recherches A\'eronautiques (ONERA), BP72 - 29 avenue de la Division Leclerc, 92322 Chatillon Cedex. France
}

\author{Michael Atlan}

\affiliation{
Institut Langevin. Fondation Pierre-Gilles de Gennes. Centre National de la Recherche Scientifique (CNRS) UMR 7587, Institut National de la Sant\'e et de la Recherche M\'edicale (INSERM) U 979, Universit\'e Pierre et Marie Curie (UPMC), Universit\'e Paris 7. Ecole Sup\'erieure de Physique et de Chimie Industrielles (ESPCI) - 1 rue Jussieu. 75005 Paris. France
}

\date{\today}

\begin{abstract}

We report on a wide-field optical monitoring method for revealing local delaminations in sandwich-type composite plates at video-rate by holographic vibrometry. Non-contact measurements of low frequency flexural waves is performed with time-averaged heterodyne holography. It enables narrowband imaging of local out-of-plane nanometric vibration amplitudes under sinusoidal excitation, and reveals delamination defects, which cause local resonances of flexural waves. The size of the defect can be estimated from the first resonance frequency of the flexural wave and the mechanical parameters of the observed layer of the composite plate.

\end{abstract}

\maketitle


Lamb waves, which interact with delaminations are useful for the inspection of laminated composites and sandwich materials \cite{SuYeLu2006, DiamantiSoutisHodgkinson2005, DiamantiSoutis2010}. Scanning laser Doppler interferometric methods were introduced and used for non-contact sensing of Lamb waves excited in a structure \cite{Audoin2002, Staszewski2004, Mallet2004}. It was shown that flaw maps can be obtained by exploiting the local vibrational contrast between a delaminated area and the surrounding structure \cite{SohnDuttaYang2011, RoggeJohnston2011}. Analysis methods of time-domain laser Doppler measurements were introduced to reveal local damage contrasts. Among them, spatial and temporal discrete Fourier transform approaches \cite{SohnDuttaYang2011, RoggeJohnston2011} and computation of the local energy from temporal wave fields \cite{LamboulPassillyRoche2013} have proven their usefulness. However, scanning laser Doppler vibrometry is a time-consuming process which hinders high spatial resolution measurements in real-time.\\

Non-destructive imaging of structural integrity can also be achieved by speckle pattern shearing interferometry (shearography)~\cite{TohShang1991, Hung1999, FrancisTatamGroves2010}, speckle interferometry~\cite{Lokberg1984}, and holography~\cite{Powell1965, Aleksoff1971, Erf1974}, which are suitable experimental methods to reveal local delaminations, especially in time-averaged recording conditions~\cite{Powell1965, Aleksoff1971, Lokberg1984, TohShang1991}. In shearing interferometry, if the image of the object is sampled on the sensor, interference fringes correspond to the partial derivative of the object's out-of-plane vibrational motion in the direction of the shear, or isoclines. In speckle interferometry and holography, an image of the local out-of-plane vibrational motion is obtained; interference fringes correspond to constant vibration amplitudes, or contour lines.\\

In the reported work, wide-field structural health monitoring of locally-damaged sandwich composite plates is performed at video rate by frequency-tunable time-averaged holographic vibrometry. Holographic vibrometry was previously reported to enable wide-field surface acoustic wave monitoring in sinusoidal regime and compared to scanning laser Doppler for this purpose~\cite{BrunoLaurentRoyer2013}. We demonstrate experimentally that delamination defects can be observed from local narrowband measurements of out-of-plane vibration amplitudes in an aluminum honeycomb core sandwich composite plate. The size of the defect is estimated from the first resonance frequencies of the flexural wave.\\


%
\begin{figure}[]
\centering
\includegraphics[width = 8.0 cm]{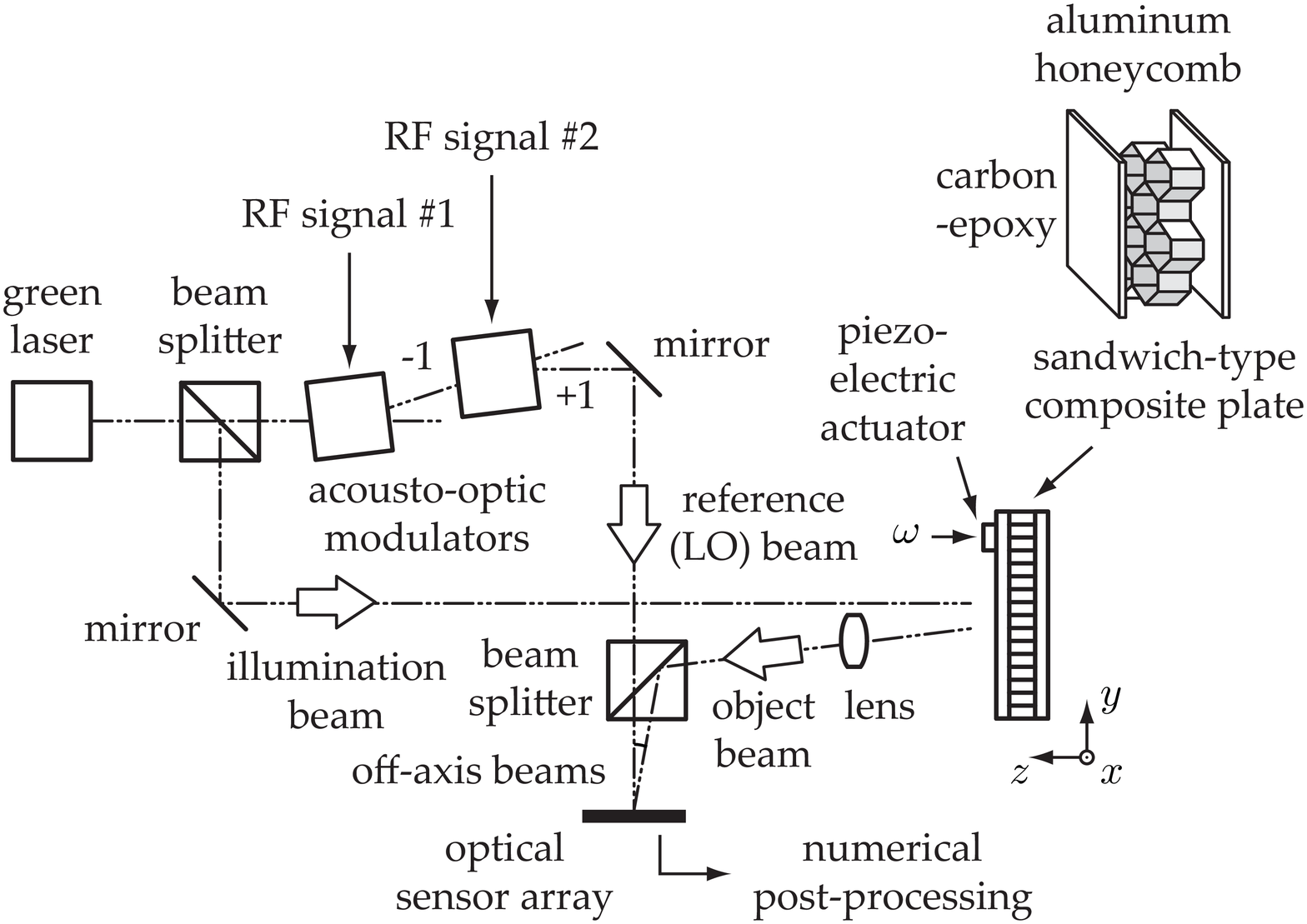}
\caption{Mach-Zehnder optical holographic interferometer. The main laser beam is split into two channels. In the object channel, the optical field is backscattered by the object in vibration at the angular frequency $\omega$, which generates two modulation sidebands at low vibration amplitudes. In the reference channel, the optical field is frequency-shifted by two acousto-optic modulators. The sensor array of the camera records the interference pattern of both optical fields beating against each other, in time-averaging conditions. Images are calculated with a standard holographic rendering algorithm involving a numerical Fresnel transform \cite{Schnars2002}.}
\label{fig_SetupHealthMonitoringOfComposites}
\end{figure} 
An optical Mach-Zehnder interferometer (Fig.~\ref{fig_SetupHealthMonitoringOfComposites}) was developed to monitor out-of-plane vibrations. Narrowband imaging is achieved with a frequency-tunable time-averaged laser Doppler holographic imaging scheme on a sensor array. This method enables robust and quantitative mapping out-of-plane vibrations of nanometric amplitudes at radiofrequencies. A high temporal coherence green laser (Cobolt Samba-TFB-150, linewidth $< 1 \, \rm MHz$, wavelength $\lambda$ = 532 nm) was used to illuminate the composite plate in wide field over $\sim$ 20 cm $\times$ 20 cm with a total optical power of $\sim$ 30 mW. This plate is excited by a piezoelectric actuator. Narrowband recording of the map of out-of-plane vibration amplitudes was enabled by holographic interferometry in time-averaging conditions by a 20 Hz frame rate CCD camera. In such experimental conditions, retrieval of nanometric amplitude oscillations is achieved. The detection process involves both spatial and temporal modulation of the interference pattern through off-axis and frequency-shifting holography \cite{AtlanGross2007JOSAA}, frequency-division multiplexing of the optical local oscillator ensures simultaneous measurement of two modulation sidebands at distinct beating frequencies of the recorded interferogram. Robust imaging of out-of-plane vibration amplitudes at a given frequency is achieved by pixel-to-pixel division of two sideband holograms \cite{VerrierAtlan2013, BrunoLaurentRoyer2013}.\\

\begin{figure}[]
\centering
\includegraphics[width = 8 cm]{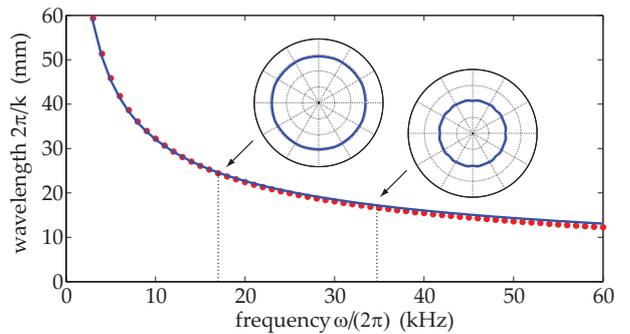}
\caption{Dispersion relation of the first antisymmetric Lamb mode ($A_0$) in a 1.1 mm-thick carbon fiber epoxy plate. The dots are derived from numerical resolution of the Rayleigh-Lamb equation \cite{BookRoyerDieulesaint2000} by the SAFE method. The line corresponds to the expression of the Rayleigh-Lamb equation in the low frequency approximation for homogeneous and isotropic plate, for which $hk \ll 1$, where $h$ is the plate's thickness and $k$ is the wave vector in Eq. \eqref{eq_DispersionRelationship}. Insert : slowness curves calculated for Lamb modes at 17 kHz and 34.5 kHz.}
\label{fig_DispersionRelationAluminiumA0}
\end{figure}

\begin{figure}[]
\centering
\includegraphics[width = 8.0 cm]{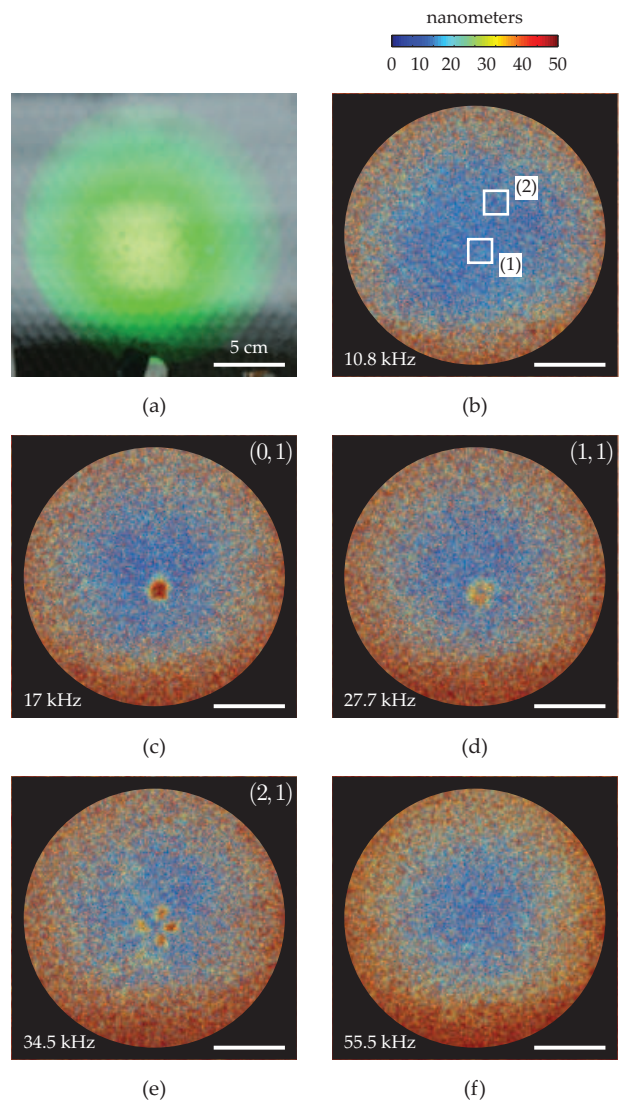}
\caption{Direct image of the composite plate used for the experiment, illuminated with the green laser (a). Holographic images of the out-of-plane vibration (b-f). Geometric vibration modes : $(m,n)= (0,1)$ at 17 kHz (c), $(m,n)= (1,1)$ at 27.7 kHz (d), $(m,n)= (2,1)$ at 34.5 kHz (e). Zone (1) corresponds to the delamination region, and zone (2) is a healthy region, 5 cm away from zone (1).}
\label{fig_CompositePlateInVibration}
\end{figure}

The composite is a symmetric sandwich, whose core is a $2.5 \, {\rm cm}$-thick aluminum honeycomb with 3/8" cells. The skin is a $1.1 \, \rm mm$-thick [0 45 90 -45]s stack of woven carbon fiber plies with epoxy resin of density $\rho = 1.7 \, \rm g/cm^3$. Its standard mechanical properties \cite{Audoin2002}, reported in Tab.~\ref{tab_rendering} for one woven ply are used to calculate the dispersion relation of the first antisymmetric Lamb mode $A_0$ by the semi-analytical finite element (SAFE) method \cite{TerrienDeom2007, LiuAchenbach1994}. This dispersion relation is plotted in Fig.~\ref{fig_DispersionRelationAluminiumA0}. The composite was impacted at 1 J on the monitoring side to provoke a local detachment of the skin from the honeycomb, invisible to the naked eye. In order to create steady-state Lamb waves in the carbon/epoxy plate, a piezo-electric transducer was fixed to the plate as shown in the bottom of Fig.~\ref{fig_CompositePlateInVibration}(a). The piezo disc ($2 \, {\rm cm}$ diameter) is supplied with a sinusoidal signal whose angular frequency $\omega$ is swept. Monitoring of resonant Lamb waves is performed by holographic vibrometry.\\ 

\begin{table}[!h]
\centering
$C =\begin{pmatrix}
60 & 15    & 7  & 0 & 0 & 0\\
   & 60    & 7  & 0 & 0 & 0\\
   &       & 14 & 0 & 0 & 0\\
   &       &    & 4 & 0 & 0\\
   \multicolumn{3}{c}{\rm sym.}  &  & 4 & 0\\
   &       &    &   &   & 7  
\end{pmatrix}$
\caption{Stiffness tensor components $C_{ij}$ of the composite material, in GPa. Plies are stacked along the out-of-plane direction $z$.}
\label{tab_rendering}
\end{table}

%
%


%

To characterize the local delamination defect, we assume that its geometry is circular and we show that the modification of the boundary conditions in the carbon fiber epoxy plate results in the creation of local resonances of anti-symmetric Lamb waves at low frequencies, which are related to the defect size. The motion of flexural $A_0$ Lamb waves in an isotropic plate is given by the Kirchhoff–Love equation
\begin{equation}\label{eq_PlateFlexuralMotion}
	\frac{\partial^2z}{\partial t^2} + \frac{Eh^2}{12\rho\left(1-\nu^2\right)}\nabla^4z = 0
\end{equation}
where $z$ is the out-of-plane motion, $\nabla$ is the gradient operator, $h$ is the thickness of the plate, $E$ is the Young modulus of the material, $\rho$ its density , and $\nu$ its Poisson's ratio. In polar coordinates $(r, \theta)$, the harmonic solutions of eq.\ref{eq_PlateFlexuralMotion} take the form $z = Z \left(r,\theta\right) \exp(i\omega t)$, where $i$ is the imaginary unit. The Rayleigh-Lamb dispersion relation of the $A_0$ mode in the low frequency approximation \cite{BookRoyerDieulesaint2000} is introduced to derive the steady-state solutions $Z(r,\theta)$ 
\begin{equation}\label{eq_DispersionRelationship}
\omega = {1 \over \sqrt{12}}k^2c_p h
\end{equation}
where $k$ is the wave vector and $c_p = \sqrt{E/{\rho\left(1-\nu^2\right)}}$ is the velocity of propagation of longitudinal oscillations in the isotropic thin plate (referred to as "plate" velocity). In our case, this expression of $c_p$ is not valid in an anisotropic material. Nevertheless, in the monitored frequency range, from 10 kHz to 60 kHz, we show that the composite can be assimilated to be a quasi-isotropic ply for which $c_p$ can be assessed from a fitting procedure of dispersion relationships. Slowness curves for Lamb modes at 17 kHz and 34.5 kHz were calculated from the values of the mechanical properties of the plate (stiffness tensor, thickness, angles between woven plies). These polar curves, which indicate a quasi-isotropic behavior of the material at these frequencies, are reported in Fig.~\ref{fig_DispersionRelationAluminiumA0}. From the values of the mechanical parameters, we derive an equivalent velocity $c_p = 5150 \, \rm  m/s$ by fitting the dispersion relation calculated from the SAFE method and reported in Fig.~\ref{fig_DispersionRelationAluminiumA0} (dots), with Eq.\ref{eq_DispersionRelationship} (plotted as a line).\\

The steady-state equation of flexural waves $\left(\nabla^2 - k^2\right)\left(\nabla^2 + k^2\right)Z = 0$ is derived from Eq.~\ref{eq_DispersionRelationship} and the harmonic solutions of Eq.~\ref{eq_PlateFlexuralMotion}. Its solutions $Z_m$, in a circular plate clamped at its borders, are linear combinations of Bessel functions multiplied by an angular function
\begin{equation}\label{eq_PlateGeneralSolution}
	Z_m = \left( A_m J_m \left( kr\right) + C_m I_m \left( kr\right) \right) \cos( m\theta + \phi )
\end{equation}
where $A_m$ and $C_m$ are constants. Here, $J_m$ are the ordinary Bessel functions and $I_m$ are the hyperbolic Bessel functions of the first kind : $I_m(kr) = i^{-m}J_m(ikr)$. For a circular plate of radius $a$, clamped at its edge $r = a$, the boundary conditions are $Z(a)= 0$ and $\partial_r Z(a) = 0$. Which provide a condition on $ka$ for solutions to be non-trivial. Using the recursive relationships of the Bessel functions, this condition takes the form $J_m(ka) I_{m+1}(ka) + I_m(ka) J_{m+1}(ka) = 0 $ where the eigenvalues $ka$ determine the resonant frequencies $\omega$ via Eq. \eqref{eq_DispersionRelationship}. The resulting allowed values are labeled 
\begin{equation}\label{eq_gamma}
\gamma_{mn} = k_{mn}a
\end{equation}
where $m$ is the number of nodal diameters and $n$ is the number of nodal circles in the corresponding normal mode \cite{Leissa1969}. Using Eq.~\ref{eq_DispersionRelationship} and Eq.~\ref{eq_gamma}, the vibration's angular frequency, $\omega_{mn} = 2 \pi f_{mn}$, is related to the radius $a$ of the delamination defect by~\cite{Morse1986}
\begin{equation}\label{eq_ftoadisperse}
	\omega_{mn} = \frac{1}{\sqrt{12}}\frac{\gamma^2_{mn}}{a^2}c_p h
\end{equation}
The first three solutions $\gamma_{mn}$ that will be used for the interpretation of the experimental results satisfy \cite{Leissa1969} $\gamma_{01} = \sqrt{10.22}$, $\gamma_{11} = \sqrt{21.26}$, and $\gamma_{21} = \sqrt{34.88}$.\\


%
\begin{figure}[]
\centering
\includegraphics[width = 8.0 cm]{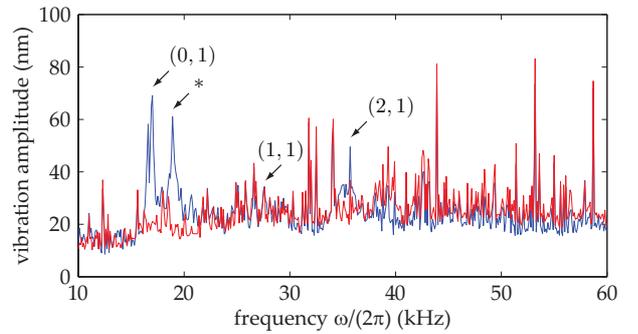}
\caption{Local out-of-plane vibration amplitude versus excitation (and measurement) frequency $\omega/(2\pi)$ in the delamination region of the composite plate (blue line, region (1)), and 5 cm away from the region (red line, region (2)). Specific resonances of the delamination region corresponding to the spatial vibration modes $(m,n)= (0,1)$, $(m,n)= (1,1)$, and $(m,n)= (2,1)$ are highlighted by arrows. The peak highlighted by $*$ is an alternate $(m,n)= (0,1)$ mode assumed to be caused either by (i) anisotropy in the carbon epoxy plate and/or (ii) non-circular symmetry of the delamination defect.}
\label{fig_AmplitudeVersusFrequency}
\end{figure} 

\begin{figure}[]
\centering
\includegraphics[width = 8.0 cm]{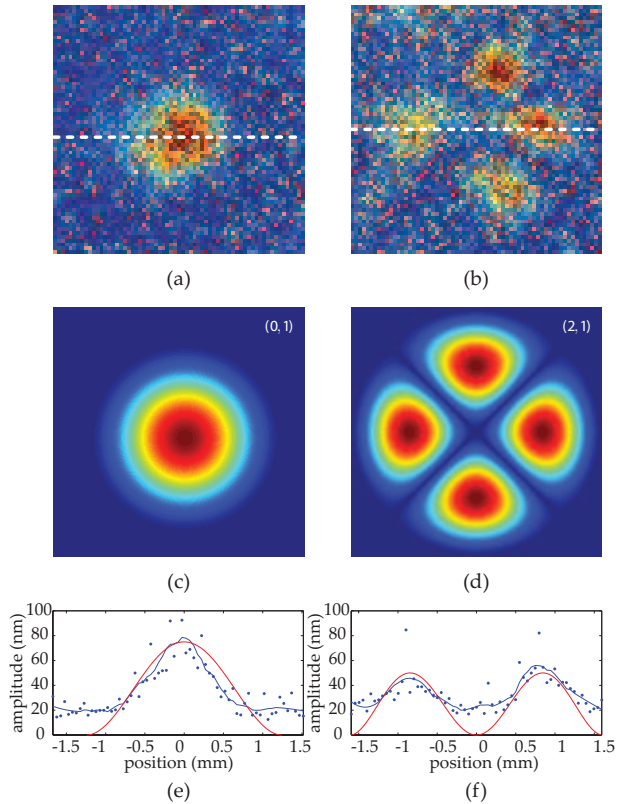}
\caption{Lateral extension of the vibrating region for the spatial vibration modes $(m,n)= (0,1)$ at 17 kHz and $(m,n)= (2,1)$ at 34.5 kHz. Holographic images of experimental results (a,b). Calculated modes (c,d). Profile of the out-of-plane amplitude along a line (e,f); blue : experiments, red : theoretical curve.}
\label{fig_ImageAndProfile}
\end{figure}

Holographic images of the amplitude of the out-of-plane vibration at different frequencies are shown in Fig.~\ref{fig_CompositePlateInVibration}. A frequency scan of the narrowband detection was performed from 10 kHz to 60 kHz for vibration monitoring. The total acquisition time for each image was 0.5 s. We assumed visually that (i) Fig.~\ref{fig_CompositePlateInVibration}(c) corresponds to the vibration mode $(m,n) = (0,1)$ of the delamination defect, (ii) Fig.~\ref{fig_CompositePlateInVibration}(d) corresponds to the mode $(m,n) = (1,1)$, and (iii) Fig.~\ref{fig_CompositePlateInVibration}(e) corresponds to the mode $(m,n) = (2,1)$. Two regions, (1) and (2), were defined in Fig.~\ref{fig_CompositePlateInVibration}(b), in the delamination region and in a typical healthy part respectively. Vibration amplitude spectra in these two zones are compared in Fig.~\ref{fig_AmplitudeVersusFrequency}. The observed resonance frequencies of the geometrical modes (0,1), (1,1), and (2,1) allowed us to estimate the value of the radius $a$ of the damaged region, derived from Eq. \ref{eq_ftoadisperse}. We obtained $a_{01} = 12.5 \, {\rm mm}$, $\ a_{11} = 14.1 \, {\rm mm}$, and $a_{21} = 16.2 \, {\rm mm}$, from the known parameters $h$ and $c_p$. These radius values are of the same order of magnitude. To validate this theoretical approach of delamination vibrations, we compared the lateral extension of the vibrating region for the mode $(m,n) = (0,1)$ and mode $(m,n) = (2,1)$, which exhibit a good signal-to-noise ratio, contrary to the mode $(m,n) = (1,1)$ (Fig.~\ref{fig_AmplitudeVersusFrequency}). In Figure \ref{fig_ImageAndProfile}, we reported images of the local out-of-plane vibration amplitude measured, respectively on (a) and (b), and of the simulated solutions of the steady-state modes $Z$, for the eigenvalues $\gamma_{01}$ (c) and $\gamma_{21}$ (d), calculated from Eq. \ref{eq_PlateGeneralSolution}. Both coefficients $A_m$ and $C_m$ for each mode were estimated using the boundary condition $Z(a_{mn}) = 0$ and the maximum amplitude measured on the experimental data. Figures~\ref{fig_ImageAndProfile} (e) and (f) show measured profiles of the vibration amplitude against calulated ones, for which good agreement is observed.\\

%


In conclusion, we have presented a real-time and wide-field structural health monitoring method for revealing delamination defects in composite plates of sandwich type. Low frequency flexural waves were generated in the plate with an ultrasound actuator and non-contact vibrometric measurements were performed with time-averaged heterodyne holography with a dual local oscillator. This scheme was used for quantitative vibration monitoring by sampling of two optical modulation sidebands within the camera bandwidth. Coupling of flexural waves with delamination defects were observed at the excitation frequency in an aluminum honeycomb core sandwich composite plate. Vibration spectra inside and away from the defect were measured by sweeping the excitation and detection frequencies. Local delaminations appeared to be the cause of the presence of local resonances at low frequencies in a the composite plate. The size of the defect was estimated from the first resonance frequency of the flexural wave and the mechanical parameters of the observed layer of the plate. Experimental results agree well with the reported analytical relationship between the vibration frequency of a local delamination of the composite and its lateral extension.\\


We gratefully acknowledge support from Fondation Pierre-Gilles de Gennes (FPGG014), Agence Nationale de la Recherche (ANR-09-JCJC-0113, ANR-11-EMMA-046), r\'egion \^Ile-de-France (C'Nano, AIMA).


\end{document}